\documentclass[twocolumn,floatfix]{revtex4}%
\usepackage[dvipdfmx]{graphicx}
\usepackage[dvipdfmx]{color}
\usepackage{amsmath,amssymb,amsfonts}
\usepackage{mathrsfs}
\usepackage{mathtools}
\usepackage{here}

\begin{document}
\title{
Giant impurity effects on charge loop current order states in kagome metals
}
\author{
Seigo Nakazawa,$^1$ Rina Tazai,$^2$ Youichi Yamakawa,$^1$ Seiichiro Onari,$^1$ and Hiroshi Kontani$^1$
}
\date{\today }

\begin{abstract}
The exotic electronic states in the charge loop current (cLC) phase, 
in which the permanent charge current breaks the time-reversal symmetry,
have been attracting increasing attention
in recently discovered kagome metals $A$V$_3$Sb$_5$ ($A$ = Cs, Rb, K).
Interestingly, the cLC state is sensitively controlled by 
applying a small magnetic field as well as a tiny uniaxial strain.
In addition, many experiments indicate that the cLC state is 
sensitive to the small number of impurities.
To understand the impurity effects on the cLC electronic states accurately,
we analyze the giant unit-cell (up to 1200 sites) kagome lattice model
with single impurity potential.
The loop current is found to be strongly suppressed 
within the current correlation length $\xi_J$ centered on the impurity site, 
where $\xi_J$ increases as the cLC order parameter $\eta$ decreases.
(The cLC order is the pure imaginary hopping integral modulation $\delta t_{i,j}=\pm i\eta$.)
In addition, both the uniform orbital magnetization $M_{\rm orb}$ and the anomalous Hall conductivity $\sigma_{xy}$ are drastically suppressed by dilute impurities.
Especially, the suppression ratio
$R=-\Delta M_{\rm orb}/M_{\rm orb}^0$
can exceed $50\%$ with the introduction of $1\%$ impurities.
Unexpectedly, the ratio $R$ is qualitatively insensitive to $\eta$, in highly contrast to a naive expectation that $R$ is proportional to the current suppression area $\pi \xi_J^2$.
The resulting giant impurity effect of $M_{\rm orb}$ would originates from the nonlocal contribution of the itinerant circulation of electrons.
The present study gives a natural explanation of why
the cLC electronic states in kagome metals
are sensitive to dilute impurities.

\end{abstract}

\affiliation{
$^1$Department of Physics, Nagoya University,
Furo-cho, Nagoya 464-8602, Japan\\
$^2$Yukawa Institute for Theoretical Physics, Kyoto University, Kyoto 606-8502, Japan.
}
\sloppy

\maketitle

\section{Introduction}
Recently, various quantum phase transitions have been reported 
in strongly correlated electron systems.
Among them, kagome metals $A$V$_3$Sb$_5$ ($A$ = Cs, Rb, K) \cite{Ortiz1,Ortiz2} 
have various interesting phases such as the charge density wave (CDW) 
\cite{BO1,BO2,BO3,BO4}, the nematic state 
\cite{BO2,BO3,nematicity2,torque,Tazai1}, and superconductivity 
\cite{SC1,SC2}.
Thus, they have attracted even more attention.
The CDW originats from the $2\times2$ Star-of-David or trihexagonal bond order 
(BO) \cite{Tazai1}. 
Significantly, the time-reversal symmetry (TRS) breaking phase without any spin polarization 
has been reported with muon spin rotation and relaxation method \cite{mSR1,mSR2,mSR3,mSR4}, NMR \cite{NMR}, 
scanning tunneling microscopy (STM) \cite{BO1}, the Kerr effect \cite{nematicity2,Kerr2}, and 
the observation of anomalous Hall conductivity \cite{AHE1,AHE2,AHE3}.
Especially, recent magnetic torque measurements \cite{torque} and 
electronic magnetochiral anisotropy \cite{eMChA} 
have presented very convincing evidence of the TRS breaking phase,
although its onset temperature $T_{\rm cLC}$ is still under debate.
Interestingly,
Ref. \cite{strain} recently reported that $T_{\rm cLC}\approx0$
in an ideal strain-free system, while $T_{\rm cLC}$ drastically increases
under weak magnetic fields or tiny strains.
Consistently, it has theoretically been found that
the TRS breaking state is sensitively controlled by 
applying a small magnetic field as well as a tiny uniaxial strain
\cite{Tazai3}.

The permanent charge loop current (cLC) state without any spin polarization
is a natural candidate for the origin 
of the TRS breaking phase \cite{Balents,cLC}.
The cLC order is the ``purely imaginary'' modulation in the hopping integral, 
$\delta t^{\mathrm{C}}=\pm i\eta$, whereas the BO with the TRS preserved 
is the ``real'' one, 
$\delta t^{\mathrm{B}}=\pm\phi$ 
\cite{BO_theory1,BO_theory2,BO_theory3,BO_theory4,Balents,Tazai1,Fernandes}.
Theoretically, the cLC is induced by the BO fluctuations \cite{Tazai2} in kagome metals.
The actual current distribution $J$ in real space is not simply proportional 
to the cLC order parameter $\eta$, as discussed in Ref. \cite{Shimura}.
Therefore, $J$ needs to be calculated microscopically.
In the triple-$\boldsymbol{Q}$ cLC state,
where the current flows in all three directions in the two-dimensional model,
the global TRS is broken, and therefore, the chirality ($\chi_{\rm ch}=\pm1$) is well-defined.
($\chi_{\rm ch}$ changes sign under the time-reversal operation, while 
$\chi_{\rm ch}$ is unchanged under the translational and rotational shifts.)
Therefore, the triple-$\boldsymbol{Q}$ cLC order
causes finite uniform orbital magnetization $M_{\mathrm{orb}}$ 
\cite{Balents,Tazai3}.
On the other hand, the single-$\boldsymbol{Q}$ cLC order breaks 
the local TRS, but it preserves the global TRS.
Thus, the uniform $M_{\mathrm{orb}}$ vanishes in the single-$\boldsymbol{Q}$ cLC order.

In kagome metals $A$V$_3$Sb$_5$ ($A$ = Cs, Rb, K),
the exotic electronic states in the cLC phase, in which the permanent charge current breaks the TRS, have been attracting increasing attention.
Interestingly, the cLC state is sensitively controlled by 
applying a small magnetic field as well as a tiny uniaxial strain
\cite{Tazai3,strain}.
In addition, many experiments indicate that the cLC state is 
sensitive to the small number of impurities.

For example, the observation of the quasiparticle-interference (QPI) signal 
in the STM measurements is one of the experiments demonstrating that the loop current is 
sensitive to dilute impurities.
Since the work by Jiang \textit{et al.,} it has generally been believed that QPI chirality is 
related to loop-current chirality \cite{BO1}.
This belief is experimentally supported by the observation 
that reversing the out-of-plane magnetic field leads to a reversal of QPI chirality.
Reference \cite{BO1} demonstrate that while QPI exhibits clear chirality 
in regions with low impurity concentrations [$n_{\mathrm{imp}}\sim O(10^{-3})$], 
it lacks distinct chirality and shows a weaker response to the magnetic field 
in impurity-rich regions.
Similarly, Refs. \cite{QPI,QPI2} reported that QPI signals exhibit chirality 
in regions with low impurity concentrations, 
whereas in impurity-rich regions, the QPI signal becomes nematic.
%For example, in the STM measurements, the chirality in the quasiparticle-interference 
%(QPI) signal, which is believed to originate from the loop current, 
%is observed only in the low-impurity region $(n_{\mathrm{imp}}\sim O(10^{-3}))$, 
%as reported in Refs. \cite{STM,QPI}.

Thus, it is important to study the impurity effect in the cLC order of kagome metals, 
although systematic understanding of it has not been achieved.

In this paper, we analyze the giant unit-cell (up to 1200 sites) kagome lattice model with single impurity potential to understand the impurity effects on the cLC electronic states accurately.
The loop current is found to be strongly suppressed 
within the current correlation length $\xi_J$ centered on the impurity site, 
where $\xi_J$ increases as the cLC order parameter $\eta$ decreases.
In addition, both the uniform orbital magnetization $M_{\rm orb}$ and the anomalous Hall conductivity $\sigma_{xy}$ are drastically suppressed by dilute impurities.
Especially, the suppression ratio
$R=-\Delta M_{\rm orb}/M_{\rm orb}^0$
can exceed $50\%$ with the introduction of $1\%$ impurities.
Unexpectedly, the ratio $R$ is qualitatively insensitive to $\eta$,
in highly contrast to a naive expectation that $R$ is proportional to the current suppression area $\pi \xi_J^2$ that decreases with $|\eta|$.
The resulting giant impurity effect of $M_{\rm orb}$ originates from the nonlocal contribution of the itinerant circulation of electrons.
The present study gives a natural explanation of why
the cLC electronic states in kagome metals
are sensitive to dilute impurities.

The microscopic mechanism of multiple exotic quantum phase transitions has been studied very actively. 
Mean-field theories as well as renormalization group theories based on (extended) Hubbard models were discussed in Refs. 
\cite{BO_theory3,BO_theory4,Balents,Tazai1,Tazai2,Thomale,Sushkov}.
We discovered that the Star-of-David BO state is driven by the paramagnon-interference mechanism \cite{Tazai1}, 
which is described by the Aslamazov-Larkin vertex corrections that are dropped in the mean-field level approximations.
Importantly, the BO fluctuations mediate not only $s$-wave or $p$-wave superconductivity but also the TRS breaking cLC order
\cite{Tazai1,Tazai2}. 

\section{Giant Unit-Cell Model}
\begin{figure}[htb]
    \centering
    \includegraphics[width=85mm]{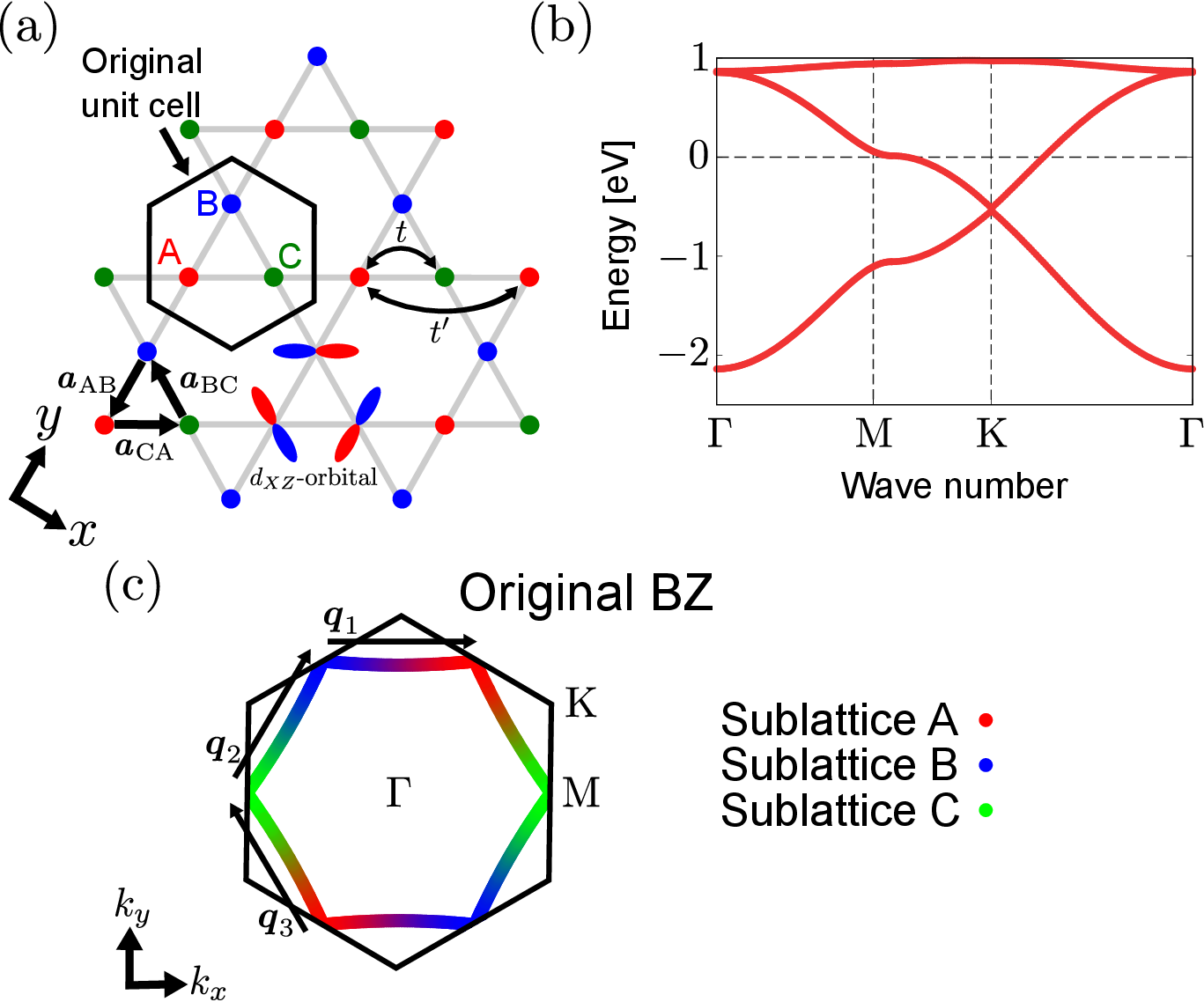}
    \caption{(a) Kagome lattice tight-binding model.
    The unit cell is composed of sublattices $A$, $B$, and $C$.
    $2\boldsymbol{a}_{\mathrm{AB}}$ and $2\boldsymbol{a}_{\mathrm{BC}}$
     give the two primitive vectors.
    The relation $\boldsymbol{a}_{\mathrm{CA}}=-\boldsymbol{a}_{\mathrm{AB}}-\boldsymbol{a}_{\mathrm{BC}}$ is held. 
    (b) Band structure and (c) Fermi surface in the original Brillouin zone (BZ) 
    at $n=n_{\mathrm{vHS}}=2.55$. 
    Here, red, blue, and green represent the weights of sublattices $A$, $B$, and $C$, respectively.}
    \label{fig1}
\end{figure}

First, we introduce the kagome lattice tight-binding model \cite{Tazai1,Thomale}
shown in Fig. \ref{fig1} (a).
The original unit cell is composed of three sublattices, $A$, $B$, and $C$.
We set the nearest-neighbor hopping integral $t=-0.5\,\mathrm{eV}$.
In addition, we introduce the nearest intra-sublattice hopping integral 
$t'=-0.02\,\mathrm{eV}$ to prevent perfect nesting and 
set the temperature $T=0.01\,\mathrm{eV}$.
Hereafter, the unit of energy is eV.
Figures \ref{fig1}(b) and 1(c) show the band structure and the Fermi surface 
at the Van Hove singularity (VHS) filling, respectively, where the VHS energy 
coincides with the chemical potential $\mu$.
In Fig. \ref{fig1}(b), the chemical potential is $\mu=0$. 
The VHS filling in the present model is given by the particle number 
$n=n_{\mathrm{vHS}}=2.55$ per three-site unit cell, where the VHS appears at the $M$ point.
In the present study, we focus on the $b_{3g}$ orbitals ($=3d_{XZ}$ orbitals) of the V ion, 
which are shown in Fig. \ref{fig1}(a).
Here, we introduce the nearest intersublattice hopping integral $t$ and 
the third-nearest intrasublattice hopping integral $t'$.
Both $t$ and $t'$ are independent of the hopping direction according to the Slater-Koster theory \cite{Slater-Koster}.
These $b_{3g}$ orbitals form the “pure-type” Fermi surface shown in Fig. \ref{fig1}(c), 
where each VHS point is composed of a single sublattice \cite{BO1,pure1,pure2,pure3,pure4}.
Here, red, blue, and green denote the contributions to the Fermi surface from sublattices $A$, $B$, and $C$, respectively.
The pure-type Fermi surface hosts three VHS points, and its bandwidth is narrower, 
so it is expected to host the correlation-driven cLC order and the bond order \cite{Tazai1}.
Here, we neglect the “mixed-type” Fermi surface composed of $b_{2g}$ orbitals 
($=3d_{YZ}$ orbitals) because the electron correlation is weaker due to its wider bandwidth,
which is found to be qualitatively important for understanding large $M_{\rm orb}$ in kagome metals
\cite{Tazai3}.
The calculation incorporating the influence of $b_{2g}$ orbitals is a topic for future works.

%%%%%%%%%%%%%%%%%%%%%%%%%%%%%%%
%%%%%%%%%%%%%%%%%%%%%%%%%%%%%%%

\begin{figure}[htb]
    \centering
    \includegraphics[width=85mm]{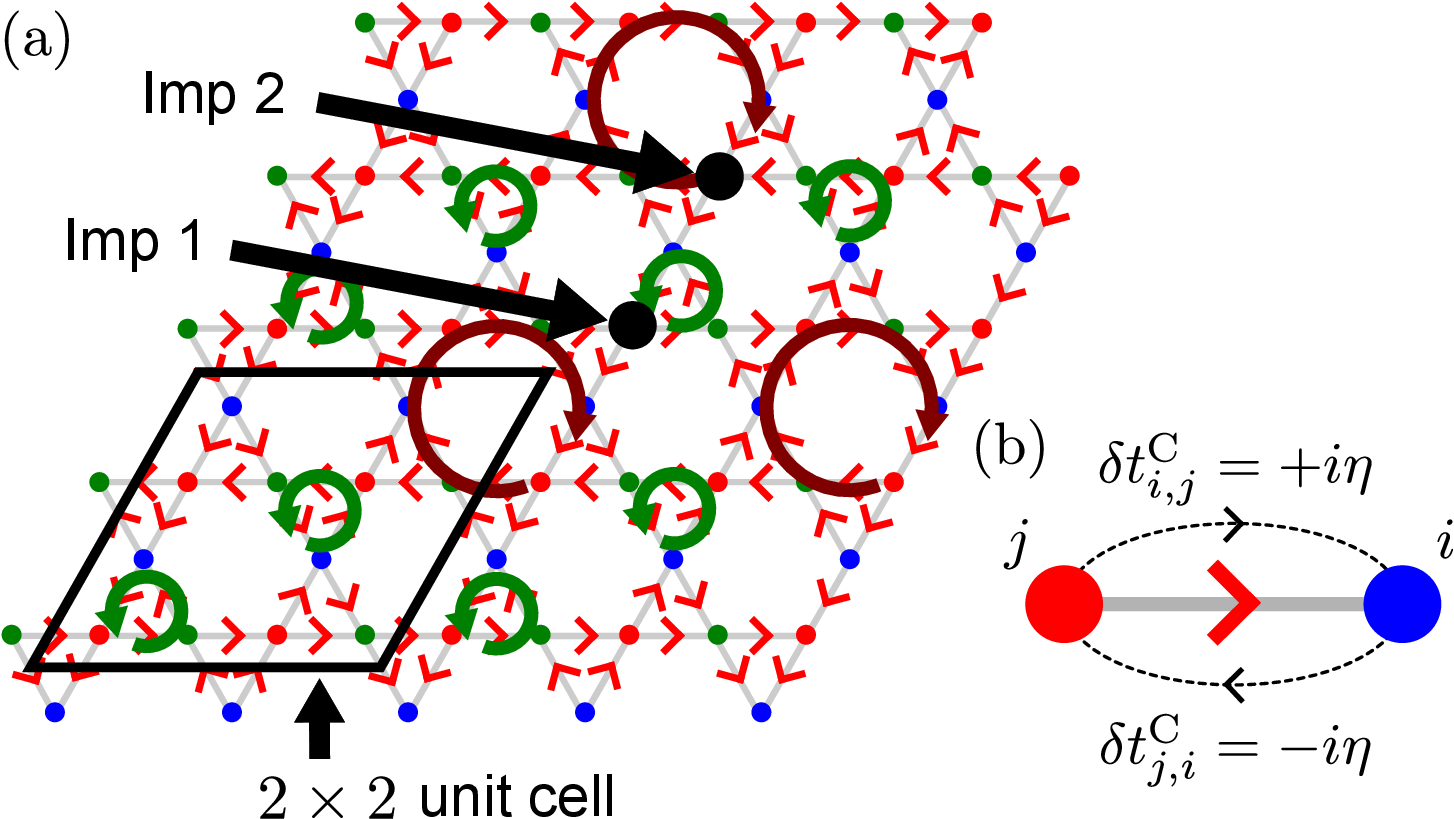}
    \caption{(a) Giant unit-cell model for $N=48$ with $M_x=M_y=2$, which corresponds to 
    $n_{\mathrm{imp}}=2.08\%$. 
    The unit cell is extended to $2\times2$, including 12 sites.
    The red arrows depict the triple-$\boldsymbol{Q}$ cLC order. 
    The black points represent the position of an introduced impurity.
    The impurity is introduced on a triangular (hexagonal) current loop 
    in the Imp 1 (Imp 2) case.
    (b) Purely imaginary and odd-parity modulation of the hopping integral that induces the cLC.}
    \label{fig2}
\end{figure}

The cLC is induced by adding the extra purely imaginary 
and odd-parity hopping integral $\delta{t}^{\mathrm{C}}=\pm{i}\eta$, 
where $\eta$ is a real number \cite{Haldane,Varma,Thomale}.
$\delta{t}^{\mathrm{C}}$ is the symmetry breaking in the self-energy with TRS breaking, which appears due to the coexistence of the strong electron correlation 
and the geometrical frustration in kagome metals \cite{Tazai2}.
Figure \ref{fig2}(a) shows the triple-$\boldsymbol{Q}$ cLC state derived from 
the density wave equation \cite{Onari,fRG,DW_equation,Jianxin} 
in a previous theoretical study \cite{Tazai2}.
In the triple-$\boldsymbol{Q}$ cLC order, the unit cell is extended to $2\times2$, which includes 12 sites.
In Fig. \ref{fig2}(a), the arrow from site $j$ to site $i$ represents 
$\delta{t}_{i,j}^{\mathrm{C}}=+i\eta$, as shown in Fig. \ref{fig2}(b).
We note that $\delta{t}_{i,j}^{\mathrm{C}}=-\delta{t}_{j,i}^{\mathrm{C}}$.
Here, we assume that $(i,j)$ is a nearest-neighbor bond for simplicity. 
Then, $|\delta{t}_{i,j}^{\mathrm{C}}|$ is equal to $\eta$ for all the nearest bonds \cite{Tazai2}. 
This cLC order corresponds to an odd parity for a site exchange operation.
For a more detailed explanation of  the cLC order in kagome metals, 
see Refs. \cite{Tazai2,Shimura}.
Reference \cite{Shimura} points out that the current $J_{i,j}$ is not proportional 
to $\delta t_{i,j}^{\rm C}$, and three different $|J_{i,j}|$ appear even when $|\delta t_{i,j}^{\rm C}|$ is the same for all nearest-neighbor bonds.
In Fig. \ref{fig2} (a), the green (brown) circular arrows correspond to the triangular (hexagonal) 
current loops.
In each circular arrow, all three or six smaller arrows have the same chirality.
Imp 1 (Imp 2) is the impurity site inserted on a triangular (hexagonal) 
loop.
Importantly, the electronic states are unchanged if Imp 1 (Imp 2) occupies distinct positions of the triangular (hexagonal) loop, if one rotates the whole system appropriately.
Because the number of triangular loops is double that of the hexagonal loops, a randomly introduced impurity belongs to Imp 1 or Imp 2 with the same probability (50\%).

To study the impurity effect in the cLC order, we construct a giant unit-cell model, 
in which the 12-site extended unit cell is arranged by $M_x\times{M}_y$, 
and introduce the single impurity potential $I$ at one of the $A$ sites.
$M_x$ and $M_y$ are integers, and we set up to $M_x=M_y=10$, 
where the total site number is $N=12M_xM_y=1200$.
Then, the corresponding impurity density is $n_{\mathrm{imp}}=0.08\%$.
We rigorously calculate the spatial distribution of the electronic states 
around the impurity without any approximation.
(For example, the position of the impurity is averaged in the $T$-matrix approximation.)
Figure \ref{fig2}(a) shows the case of $N=48$ with $M_x=M_y=2$, which corresponds to 
$n_{\mathrm{imp}}=2.08\%$.

We consider the Hamiltonian
\begin{equation}
    \label{Hamiltonian}
    \begin{split}
    H=\sum_{\boldsymbol{k},i,j,\sigma}^{\mathrm{fBZ}}h_{i,j}(\boldsymbol{k})
    c_{\boldsymbol{k},i,\sigma}^{\dagger}c_{\boldsymbol{k},j,\sigma}
    +\sum_{\boldsymbol{k},\sigma}^{\mathrm{fBZ}}Ic_{\boldsymbol{k},a,\sigma}^{\dagger}c_{\boldsymbol{k},a,\sigma},
    \end{split}
\end{equation}
where $c_{\boldsymbol{k},i,\sigma}\,(c_{\boldsymbol{k},i,\sigma}^{\dagger})$ 
is an annihilation (creation) operator of an electron for site $i$ ($i=1$-$N$), 
wave vector $\boldsymbol{k}$, and spin $\sigma$.
In Eq. \eqref{Hamiltonian}, the $\boldsymbol{k}$ summation is taken inside the 
folded Brillouin zone (fBZ), whose size is $1/4M_xM_y$ of the original BZ 
in Fig. \ref{fig1}(c).
In the first term, $h_{i,j}(\boldsymbol{k})$ is the Fourier transformation of 
the hopping integral $t_{i,j}=t^0_{i,j}+\delta t_{i,j}^{\mathrm{C}}$.
Here, $t_{i,j}^0$ denotes the original hopping integral of the kagome lattice model,
and $\delta{t}_{i,j}^{\mathrm{C}}$ is the modulation of the hopping integral 
by the cLC order.
In the second term, $I$ represents the impurity potential, and we put $I=100\,\mathrm{eV}$ in the numerical study.
This unitary limit impurity corresponds to a vacancy defect.
$a$ is the impurity site.
We set the number of $\boldsymbol{k}$ meshes $N_{\boldsymbol{k}}=512\times512$ 
in the fBZ in the numerical study.
Then, we analyze the model with $N_{\bm k}$ periodically arranged $N$-site unit cells.
That is, the number of sites included in the numerical study is $N \times N_{\boldsymbol{k}}$.

\section{Formalism}
To study the impurity effect, we calculate the current distribution 
in real space $J_{i,j}$, the uniform orbital magnetization $M_{\mathrm{orb}}$, 
and the anomalous Hall conductivity $\sigma_{xy}$.
To calculate the intersite current $J_{i,j}$ accurately by eliminating the system size effect, 
we set $N=972$ with $M_x=M_y=9$, which corresponds to $n_{\mathrm{imp}}=0.1$\%.
On the other hand, in calculating $M_{\mathrm{orb}}$ and $\sigma_{xy}$, 
we can safely set $N=300$ with $M_x=M_y=5$, 
which corresponds to $n_{\mathrm{imp}}=0.33$\%.
We have verified that the derived single impurity effects for $M_{\mathrm{orb}}$ and $\sigma_{xy}$ 
from $N=300$ model are reliable by calculating those quantities in $N=1200$ model 
(see Fig. \ref{fig8} below).
The current from site $j$ to site $i$, $J_{i,j}$, is calculated as
\begin{equation}
    J_{i,j}=i\left(t_{i,j}\left\langle c^{\dagger}_ic_j\right\rangle-t_{j,i}\left\langle c^{\dagger}_jc_i\right\rangle\right),
\end{equation}
where $\left\langle c^{\dagger}_ic_j\right\rangle$ is given by 
\begin{equation}
    \left\langle c^{\dagger}_ic_j\right\rangle=\frac{1}{N_{\boldsymbol{k}}}\sum_{n,\boldsymbol{k},\sigma}^{\mathrm{fBZ}}f_{n,\boldsymbol{k}}U_{i,n,\sigma}^*(\boldsymbol{k})U_{j,n,\sigma}(\boldsymbol{k})e^{-i\boldsymbol{k}\cdot(\boldsymbol{r}_i-\boldsymbol{r}_j)}.
\end{equation}
Here, $f_{n,\boldsymbol{k}}$ is the Fermi distribution function, 
$U_{i,n,\sigma}(\boldsymbol{k})$ denotes the unitary matrix which diagonalizes the Hamiltonian matrix, 
and $\boldsymbol{r}_i$ represents the position vector of site $i$.
The band index $n$ takes values from 1 to $N$.
We set the charge of an electron $-e$ as $-1$.
Theoretically, the magnitude of $|\eta|$ is the same for all bonds for 
$T\lesssim T_{\mathrm{LC}}$ \cite{Tazai2}. 
Nonetheless, the induced current $|J_{i,j}|$ takes three different values 
depending on the nearest bond $(i,j)$.
Next, the orbital magnetization $M_{\mathrm{orb}}$ per site can be calculated as 
\cite{Morb1,Morb2}
\begin{equation}
    \label{Morb}
    \begin{split}
    M_{\mathrm{orb}}&=\frac{1}{\pi N_{\boldsymbol{k}}N}\sum_{n,\boldsymbol{k}}^{\mathrm{fBZ}}
    \Bigl[m_{n,\boldsymbol{k}}f_{n,\boldsymbol{k}}\\
    &\left.\qquad\qquad\quad+\Omega_{n,\boldsymbol{k}}T\ln\left(1+e^{-(\epsilon_{n,\boldsymbol{k}}-\mu)/T}\right)\right],
    \end{split}
\end{equation}
where
\begin{equation}
    \begin{split}
    m_{n,\boldsymbol{k}}&=\frac{i}{2}\left[\left\langle 
        \boldsymbol{\nabla}_{\boldsymbol{k}}u_{n,\boldsymbol{k}}\middle|
        \times\left(\epsilon_{n,\boldsymbol{k}}-\hat{h}(\boldsymbol{k})\right)\middle|
        \boldsymbol{\nabla}_{\boldsymbol{k}}u_{n,\boldsymbol{k}}\right\rangle\right]_z\\
        &=\frac{1}{2}\sum_{l\ne n}\mathrm{Im}\left[\frac{\left(v_{n,l,\boldsymbol{k}}^{x}\right)^*v_{n,l,\boldsymbol{k}}^{y}-
        v_{n,l,\boldsymbol{k}}^{x}\left(v_{n,l,\boldsymbol{k}}^{y}\right)^*}
        {\epsilon_{n,\boldsymbol{k}}-\epsilon_{l,\boldsymbol{k}}}\right]
    \end{split}
\end{equation}
and
\begin{equation}
    \begin{split}
    \Omega_{n,\boldsymbol{k}}&=i\left[\left\langle\boldsymbol{\nabla}_{\boldsymbol{k}}
    u_{n,\boldsymbol{k}}\middle|\times\middle|\boldsymbol{\nabla}_{\boldsymbol{k}}
    u_{n,\boldsymbol{k}}\right\rangle\right]_z\\
    &=\sum_{l\ne n}\mathrm{Im}\left[\frac{\left(v_{n,l,\boldsymbol{k}}^{x}\right)^*v_{n,l,\boldsymbol{k}}^{y}-
    v_{n,l,\boldsymbol{k}}^{x}\left(v_{n,l,\boldsymbol{k}}^{y}\right)^*}
    {\left(\epsilon_{n,\boldsymbol{k}}-\epsilon_{l,\boldsymbol{k}}\right)^2}\right]
    \end{split}
\end{equation}
denote the orbital magnetic moment and the Berry curvature, respectively.
Here, we use the relation
\begin{equation}
    \left\langle u_{l,\boldsymbol{k}}\middle|\boldsymbol{\nabla}^{\mu}_{\boldsymbol{k}}\middle|
    u_{m,\boldsymbol{k}}\right\rangle=\frac{v_{l,m,\boldsymbol{k}}^{\mu}}
    {\epsilon_{l,\boldsymbol{k}}-\epsilon_{m,\boldsymbol{k}}},
\end{equation}
and $v_{l,m,\boldsymbol{k}}^{\mu}=\left\langle
    u_{l,\boldsymbol{k}}\middle|\boldsymbol{\nabla}^{\mu}_{\boldsymbol{k}}\hat{h}(\boldsymbol{k})
    \middle|u_{m,\boldsymbol{k}}\right\rangle$ is the matrix element of the velocity operator 
for the wave vector $\boldsymbol{k}$ in the band representation.
In eq. \eqref{Morb}, the first (second) term represents the contribution 
from the bulk (edge).
$\epsilon_{n,\boldsymbol{k}}$ denotes the energy eigenvalue, and $u_{n,\boldsymbol{k}}$ represents the Bloch wave function.

We note that $M_{\mathrm{orb}}\propto\eta^3$ without impurities, as reported 
in Ref. \cite{Tazai3} in the triple-$\boldsymbol{Q}$ cLC order shown in Fig. \ref{fig2} (a).
The sign of $M_{\mathrm{orb}}$ is the inverse when the sign of $\eta$ is the inverse.
The finite $M_{\mathrm{orb}}$ originates from the incomplete cancellation of 
the magnetic flux, which is a natural consequence of the global TRS breaking 
in the triple-$\boldsymbol{Q}$ cLC order.
In other words, the time-reversal operation of the triple-$\boldsymbol{Q}$ cLC order cannot be reproduced by its translational and rotational operations.
One may expect that $M_{\mathrm{orb}}=0$ because the magnetic fluxes 
due to the triangle loops and the hexagonal loops are opposite and cancel out.
However, this cancellation is nontrivial due to the presence of the sizable contributions 
from the distant currents, including the edge current contribution \cite{Morb1}.
Here, we reveal that $M_{\mathrm{orb}}\ne0$ in the triple-$\boldsymbol{Q}$ cLC state based on 
the exact formula for $M_{\mathrm{orb}}$ in Refs. \cite{Morb1,Morb2} 
as a natural consequence of the global TRS breaking.
On the other hand, $M_{\mathrm{orb}}$ vanishes in the single-$\boldsymbol{Q}$ 
order, in which $\eta\ne0$ only in one direction, due to the global TRS.

Furthermore, we can evaluate the anomalous Hall conductivity $\sigma_{xy}$ 
for $T=0$ as \cite{sigma}
\begin{equation}
    \sigma_{xy}=\sigma_{xy}^{\mathrm{I}}+\sigma_{xy}^{\mathrm{IIa}}+\sigma_{xy}^{\mathrm{IIb}}.
\end{equation}
Here, $\sigma_{xy}^{\mathrm{I}}$, $\sigma_{xy}^{\mathrm{IIa}}$, 
and $\sigma_{xy}^{\mathrm{IIb}}$ are given by
\begin{equation}
    \begin{split}
    \sigma_{xy}^{\mathrm{I}}=&-\frac{1}{\pi N_{\boldsymbol{k}}N}
    \sum_{l\ne m,\boldsymbol{k}}^{\mathrm{fBZ}}\mathrm{Im}\left[v_{m,l,\boldsymbol{k}}^xv_{l,m,\boldsymbol{k}}^y\right]\\
    &\times\mathrm{Im}\left[\frac{1}{\left(\epsilon_{l,\boldsymbol{k}}-\mu-i\gamma\right)
    \left(\epsilon_{m,\boldsymbol{k}}-\mu+i\gamma\right)}\right],
    \end{split}
\end{equation}
\begin{equation}
    \begin{split}
    \sigma_{xy}^{\mathrm{IIa}}=&-\frac{1}{\pi N_{\boldsymbol{k}}N}\sum_{l\ne m,\boldsymbol{k}}^{\mathrm{fBZ}}
    \mathrm{Im}\left[v_{m,l,\boldsymbol{k}}^xv_{l,m,\boldsymbol{k}}^y\right]\frac{1}{\epsilon_{l,\boldsymbol{k}}-\epsilon_{m,\boldsymbol{k}}}\\
    &\times\mathrm{Im}\left[\frac{\epsilon_{l,\boldsymbol{k}}+\epsilon_{m,\boldsymbol{k}}-2\mu-2i\gamma}{\left(\epsilon_{l,\boldsymbol{k}}-\mu-i\gamma\right)\left(\epsilon_{m,\boldsymbol{k}}-\mu-i\gamma\right)}\right],
    \end{split}
\end{equation}
and 
\begin{equation}
    \begin{split}
    \sigma_{xy}^{\mathrm{IIb}}=&\frac{2}{\pi N_{\boldsymbol{k}}N}
    \sum_{l\ne m,\boldsymbol{k}}^{\mathrm{fBZ}}\mathrm{Im}
    \left[v_{m,l,\boldsymbol{k}}^xv_{l,m,\boldsymbol{k}}^y\right]
    \frac{1}{\left(\epsilon_{l,\boldsymbol{k}}-\epsilon_{m,\boldsymbol{k}}\right)^2}\\
    &\times\mathrm{Im}\left[\ln\left(\frac{\epsilon_{l,\boldsymbol{k}}-\mu-i\gamma}{\epsilon_{m,\boldsymbol{k}}-\mu-i\gamma}\right)\right],
    \end{split}
\end{equation}
respectively.
Here, $\gamma\,(>0)$ is the imaginary part of the self-energy \cite{sigma}.
In principle, we should set $\gamma\rightarrow+0$, 
as the current $N$-site unit-cell model is periodic.
However, to achieve reliable numerical results, 
we assign a small value of $\gamma\,(0.01\,\mathrm{eV})$ to account for minor randomness 
or inelastic scattering.
$\sigma_{xy}^{\mathrm{I}}$ is the contribution 
from the Fermi surface, while $\sigma_{xy}^{\mathrm{IIa}}$ and $\sigma_{xy}^{\mathrm{IIb}}$ 
are the contributions from the Fermi sea.
Also, $\sigma_{xy}^{\mathrm{IIb}}$ represents the contribution from the Berry curvature: 
$\sigma_{xy}^{\mathrm{IIb}}\propto\sum_{n,\boldsymbol{k}}\Omega_{n,\boldsymbol{k}}f_{n,\boldsymbol{k}}$ 
for $\gamma\rightarrow0$ at $T\rightarrow0$.
We note that $\sigma_{xy}$ also vanishes 
when the global TRS is preserved.

\section{Result}
\subsection{Impurity effect in the triple-$\boldsymbol{Q}$ cLC}
First, we calculate the current distribution $J$ 
in the triple-$\boldsymbol{Q}$ cLC state with an impurity for $N=972$ with $M_x=M_y=9$, 
which corresponds to $n_{\mathrm{imp}}=0.1\,\%$.
We note that when we introduce a single impurity as a vacancy defect ($I\approx\infty$), 
the effective electron filling at other sites is modified as
\begin{equation}
    n_{\mathrm{eff}}=\frac{N}{N-1}n,
\end{equation}
where $n$ is the original electron filling.

\begin{figure}[htb]
    \centering
    \includegraphics[width=85mm]{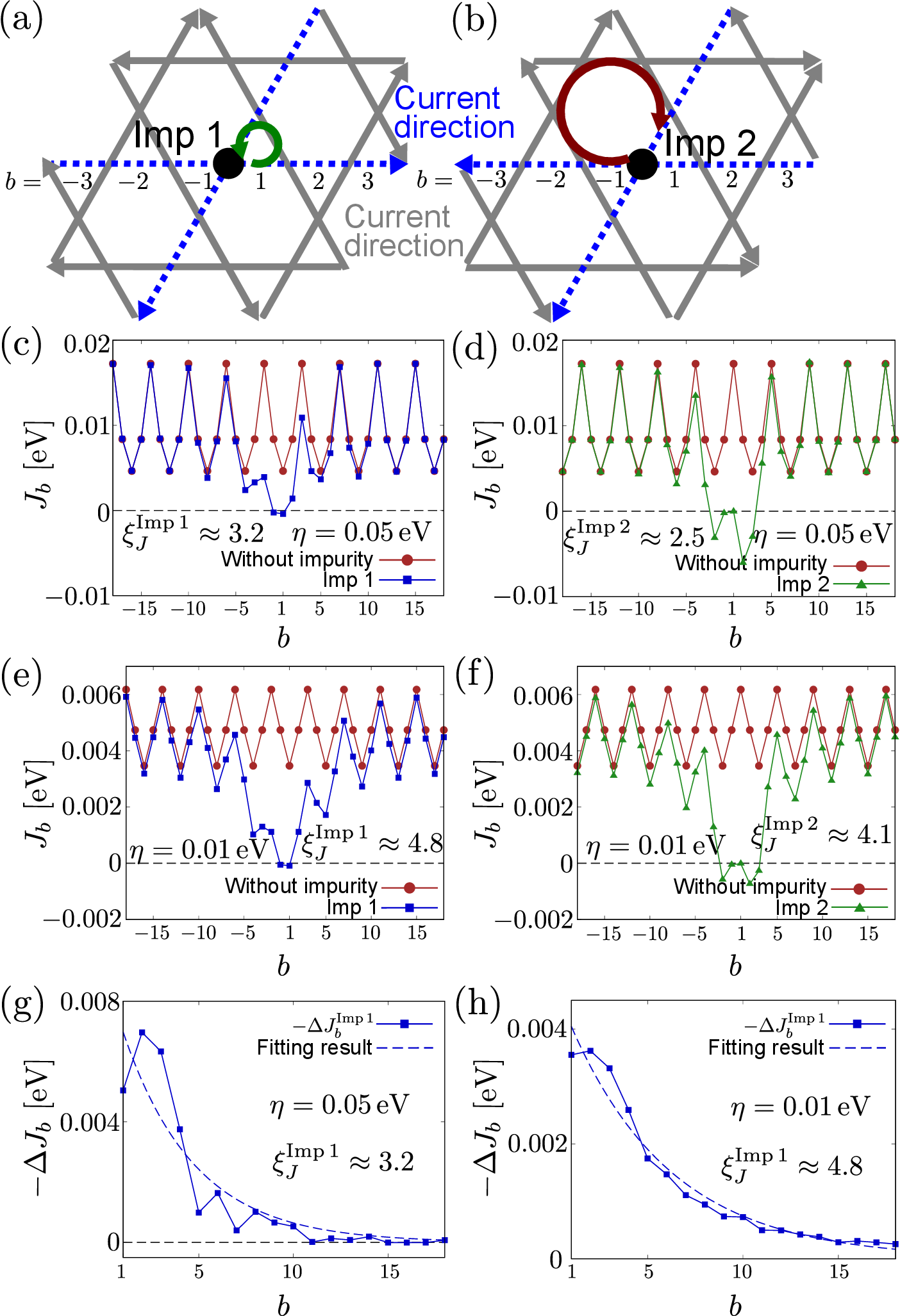}
    \caption{Schematic picture of the directions of $\eta$ and the position of the impurity 
    in the case of (a) Imp 1 and (b) Imp 2.
    The black point is the impurity site.
    The blue and gray arrows represent the current directions.
    Obtained currents on the bonds along a straight line through the impurity 
    in (c) the Imp 1 case and (d) the Imp 2 case for $\eta=0.05\,\mathrm{eV}$ and 
    $n_{\mathrm{eff}}=2.48$.
    Obtained currents on the bonds along a straight line through the impurity 
    in (e) the Imp 1 case and (f) the Imp 2 case for $\eta=0.01\,\mathrm{eV}$ and 
    $n_{\mathrm{eff}}=2.48$.
    $J_b$ is the current along the $b$th bond.
    The fitting results for (g) $\eta=0.05$ eV 
    and (h) $\eta=0.01$ eV in the case of Imp 1.}
    \label{fig3}
\end{figure}

In Figs. \ref{fig3}(a) and 3(b), the black point represents the impurity: 
Imp 1 in Fig. 3(a) and Imp 2 in Fig. 3(b).
The blue dotted arrows denote the currents along a straight line through the impurity, 
which is expected to be suppressed near the impurity site.
Other currents are shown by the gray arrows.

%%%%%%%%%%%%%%%%%%%%%%%%%%%%%%%%%%%%%%%%
%%%%%%%%%%%%%%%%%%%%%%%%%%%%%%%%%%%%%%%%

Figure \ref{fig3}(c) [Fig. 3(d)] shows the obtained currents $J_b$ along the $b$th bond 
from the impurity site on the blue dotted line in Fig. \ref{fig3}(a) [Fig. 3(b)]
for $n_{\mathrm{eff}}=2.48$.
Here, the bond index $b$ is shown in Figs.\ref{fig3} (a) and 3(b).
The obtained currents on the two blue dotted lines are the same.
These results indicate that currents near the impurity are drastically changed, 
while currents are almost unchanged by the impurity for $|b|>\xi_J$, 
where $\xi_J$ is the current correlation length.
We use the least-squares fitting method to obtain $\xi_J$.
The fitting function is 
\begin{equation}
    \left|\Delta J_b\right|=J_0\exp\left(-\frac{|b|}{\xi_J}\right),
\end{equation}
where $\Delta J_b$ is the change in the current by the impurity and 
$J_0$ is a constant value.
Because the current is in two directions, the fitting is performed for both directions, 
and we average the two current correlation lengths.
For $\eta=0.05$ eV, $\xi_J^{\mathrm{Imp}\,1}\approx3.2$ in the Imp 1 case, 
while $\xi_J^{\mathrm{Imp}\,2}\approx2.5$ in the Imp 2 case, as shown in Figs. \ref{fig3}(c) and 3(d).
For $\eta=0.01$ eV, the correlation length increases as $\xi_J^{\mathrm{Imp}\,1}\approx4.8$, 
while $\xi_J^{\mathrm{Imp}\,2}\approx4.1$, 
as shown in Figs. \ref{fig3}(e) and 3(f).
Thus, we find that $\xi_J$ increases as $\eta$ decreases.
In the Imp 1 case, the fitting results for $\eta=0.05$ eV and $\eta=0.01$ eV 
are shown in Figs. 3(g) and 3(h), respectively.
This result reminds us of the BCS coherence length 
$\xi_{\rm BCS}=\pi v_{\rm F}/\Delta_{\rm SC}$.
Under the triple-$\boldsymbol{Q}$ cLC order, 
the folded Fermi surfaces meet around the $\Gamma$ point in the folded BZ.
The band-hybridization gap $\Delta_{\mathrm{cLC}}$ due to the cLC order will be proportional to $\eta$.
Therefore, the relationship in which $\xi_J$ increases as $|\eta|\rightarrow0$
is naturally expected in kagome metals.

%%%%%%%%%%%%%%%%%%%%%%%%%%%%%%%%%%%%%%%%
\begin{figure}[htb]
    \centering
    \includegraphics[width=85mm]{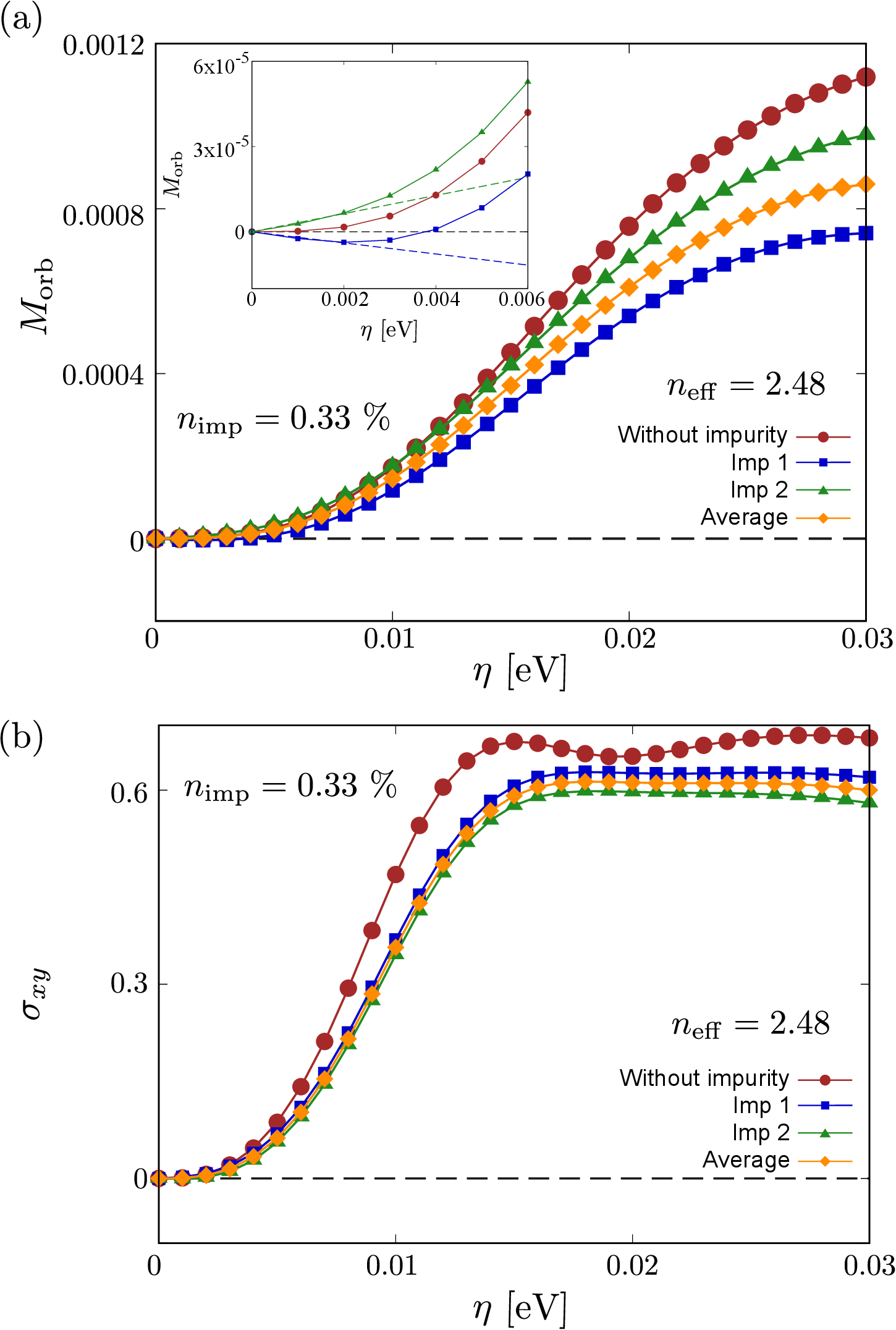}
    \caption{Obtained $\eta$ dependences of (a) $M_{\mathrm{orb}}$ and 
    (b) $\sigma_{xy}$ at $n_{\mathrm{eff}}=2.48$.
    The inset shows the obtained $M_{\mathrm{orb}}$ for $0\le\eta\le0.006$.
    The blue and green dashed lines represent the lines proportional to $\eta$.}
    \label{fig4}
\end{figure}
%%%%%%%%%%%%%%%%%%%%%%%%%%%%%%%%%%%%%%%%

Figure \ref{fig4} represents the obtained $\eta$ dependences of 
$M_{\mathrm{orb}}$ and $\sigma_{xy}$ 
in the triple-$\boldsymbol{Q}$ cLC order with an impurity for $N=300$ 
with $M_x=M_y=5$, which corresponds to $n_{\mathrm{imp}}=0.33\,\%$.
Here, $M_{\mathrm{orb}}=1$ (per V site) corresponds to 1$\mu_{\mathrm{B}}$, 
where $\mu_{\mathrm{B}}$ represents the Bohr magneton. 
Also, $\sigma_{xy}=1$ corresponds to $8.8\times10^3\,\mathrm{\Omega^{-1}cm^{-1}}$.
Hereafter, we use the same units for $M_{\mathrm{orb}}$ and $\sigma_{xy}$.
In the absence of the impurity, $M_{\mathrm{orb}}\propto\eta^3$, 
as discussed in Ref. \cite{Tazai3}.
Interestingly, we find that $M_{\mathrm{orb}}\propto\eta^1$ in both the Imp 1 and Imp 2 cases, 
as shown in the inset of Fig. \ref{fig4}(a).

For general order parameter $\boldsymbol{\eta}=(\eta_1,\eta_2,\eta_3)$, 
$M_{\mathrm{orb}}(\boldsymbol{\eta})$ is expanded as $\sum_{pqr}b_{pqr}(\eta_1)^p(\eta_2)^q(\eta_3)^r$, 
with $p+q+r=$ odd because $M_{\mathrm{orb}}(\boldsymbol{\eta})$ is an odd function of $\boldsymbol{\eta}$.
[$\eta_m$ represents the order parameter with wave vector $\boldsymbol{q}_m$; see Fig. \ref{fig1}(c).]
Also, in the case without impurities, $b_{pqr}$ vanishes unless 
$p\boldsymbol{q}_1+q\boldsymbol{q}_2+r\boldsymbol{q}_3=\boldsymbol{0}$, 
as a consequence of the momentum conversation law.
Thus, $b_{100}$, $b_{010}$, and $b_{001}$ vanish, and therefore, $M_{\mathrm{orb}}\propto\eta^3$ 
in the low-$\eta$ region \cite{Tazai3}.
However, in the case with impurities, $b_{100}$, $b_{010}$, and $b_{001}$ 
can be nonzero due to the violation of the momentum conversation law by the impurity scattering.
Therefore, $M_{\mathrm{orb}}\propto\eta$ in the low-$\eta$ region.
The orange lines represent the impurity-averaged result, 
in which the relation $M_{\mathrm{orb}}\propto\eta^3$ is recovered.
We discover that both $M_{\rm orb}$ and $\sigma_{xy}$ are drastically suppressed by dilute $(0.33\%)$ impurities.
Overall, the impurity effect on $M_{\mathrm{orb}}$ is greater 
than that on $\sigma_{xy}$.

%%%%%%%%%%%%%%%%%%%%%%%%%%%%%%%%%%%%%%%%%%%
\begin{figure}[htb]
    \centering
    \includegraphics[width=85mm]{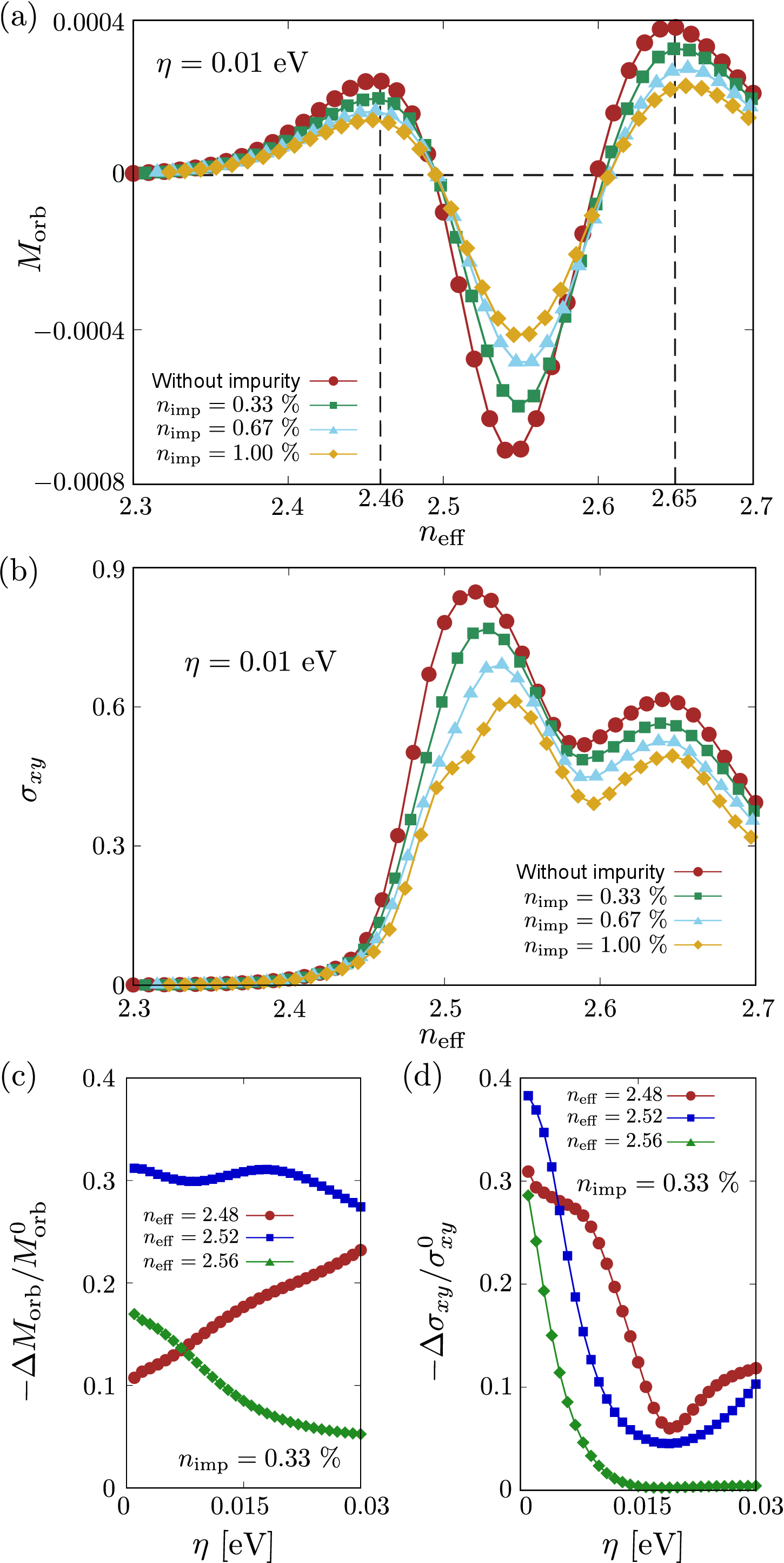}
    \caption{Obtained $n_{\mathrm{eff}}$ dependences of (a) $M_{\mathrm{orb}}$ 
    and (b) $\sigma_{xy}$ for $\eta=0.01\,\mathrm{eV}$.
    The results for $n_{\mathrm{imp}}=0.67\%,\,1.00\%$ are obtained by linear
    extrapolation of the results for $n_{\mathrm{imp}}=0,\,0.33\,\%$.
    Obtained suppression ratio of (c) $M_{\mathrm{orb}}$ and (d) $\sigma_{xy}$ 
    at $n_{\mathrm{eff}}=$ 2.48, 2.52, and 2.56.}
    \label{fig5}
\end{figure}
%%%%%%%%%%%%%%%%%%%%%%%%%%%%%%%%%%%%%%%%%%%

Figures \ref{fig5}(a) and 5(b) show the effective electron filling 
$n_{\mathrm{eff}}$ dependences of $M_{\mathrm{orb}}$ and $\sigma_{xy}$ 
under the condition of $\eta=0.01\,\mathrm{eV}$, respectively.
The results for $n_{\mathrm{imp}}=0.67\%,\,1.00\%$ are obtained by linear
extrapolation of the results for $n_{\mathrm{imp}}=0\%$ and $0.33\%$ since 
the reduction of $M_{\mathrm{orb}}$ and $\sigma_{xy}$ is proportional to
$n_{\mathrm{imp}}$ in the low impurity density region.
$|M_{\mathrm{orb}}|$ has the main peak at 
$n_{\mathrm{eff}}=n_{\mathrm{vHS}}=2.55$.
In addition, two satellite peaks appear at 
$n_{\mathrm{eff}}=2.46, 2.65$, where the condition 
$\left|\mu-\mu_{\mathrm{vHS}}\right|\sim\mathrm{max}\left\{2\eta,T\right\}$ 
is fulfilled \cite{Tazai3,Balents}.
Here, $\mu_{\mathrm{vHS}}$ is the chemical potential 
at $n=n_{\mathrm{vHS}}=2.55$.
In the present case, we set $\eta=T=0.01\,\mathrm{eV}$, so that 
$\left|\mu-\mu_{\mathrm{vHS}}\right|\sim0.02\,\mathrm{eV}$ 
is satisfied at these fillings.

Surprisingly, only $1\%$ impurities reduce uniform $|M_{\mathrm{orb}}|$ 
by approximately 40\% at arbitrary filling.
To find the impurity effects in more detail, we show the suppression ratio $R=-\Delta{X}/X^0$ at $n_{\rm imp}=0.33\%$ 
for $X=M_{\mathrm{orb}}$ in Fig. \ref{fig5}(c) and $X=\sigma_{xy}$ in Fig. \ref{fig5}(d).
Here, $\Delta X$ denotes the change given by the impurity, and $X^0$ is the quantity without the impurity.
Both $M_{\mathrm{orb}}$ and $\sigma_{xy}$ exhibit drastic suppression ratios for $\eta<0.01$.
Unexpectedly, the ratio $R$ for $M_{\mathrm{orb}}$ is qualitatively insensitive to $\eta$, 
in high contrast to the naive expectation $R\propto \xi_J^2$.
The resulting giant impurity effect of $M_{\rm orb}$ originates from the nonlocal contribution from the itinerant circulation of electrons.

\subsection{Single-$\boldsymbol{Q}$ cLC}
\begin{figure}[htb]
    \centering
    \includegraphics[width=85mm]{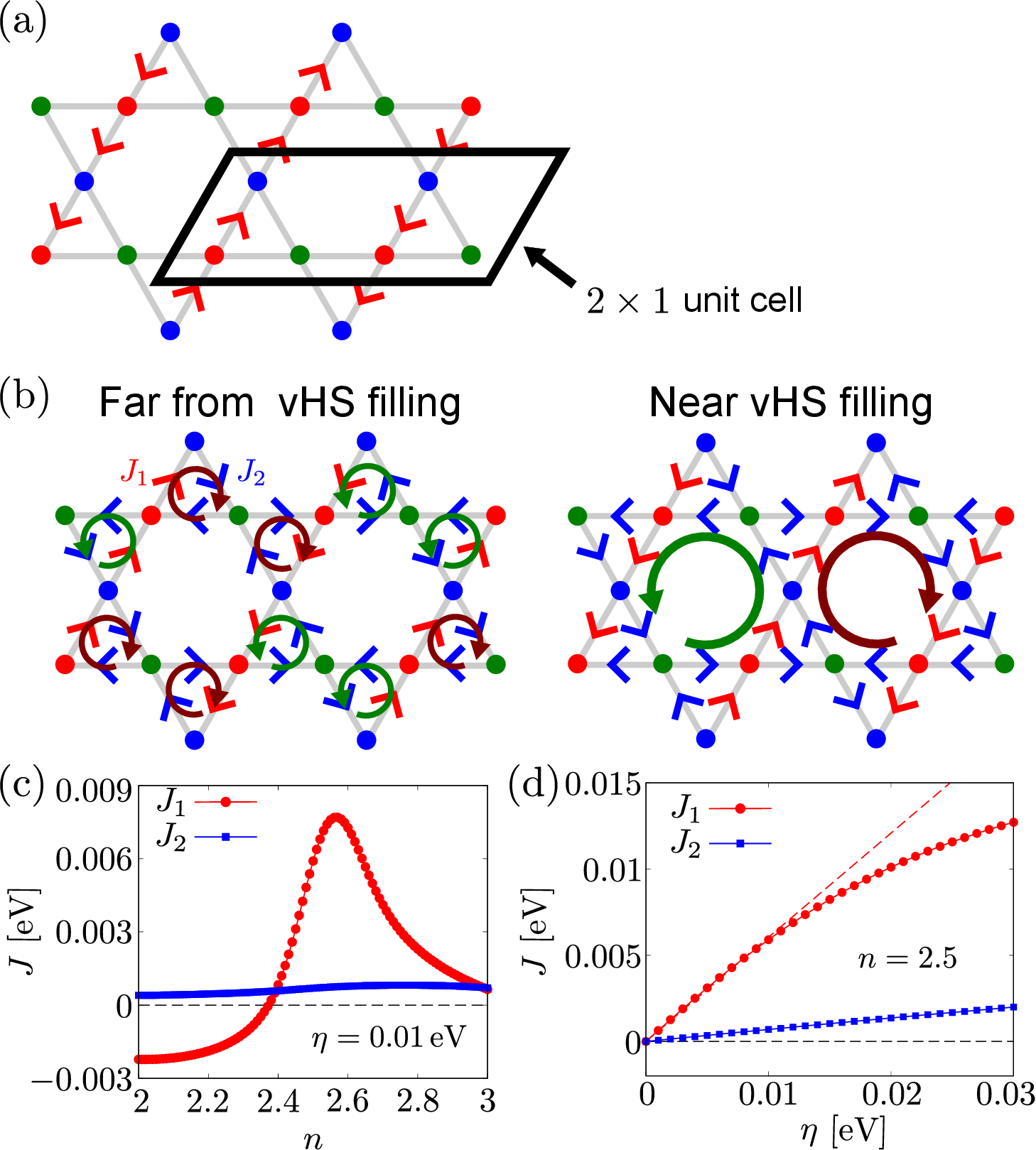}
    \caption{(a) Kagome lattice tight-binding model for 
    the single-$\boldsymbol{Q}$ cLC order. 
    The unit cell is extended to $2\times1$ in the absence of the impurity.
    The arrows represent the directions of $\eta$.
    (b) Obtained current pattern when $n$ is far from the VHS filling $n_{\rm VHS}$ (left) and $n\approx n_{\rm vHS}$ (right).
    The arrows depict the obtained current directions.
    The red arrows represent $J_1$, which flows on the $\eta\ne0$ bonds.
    The blue arrows represent $J_2$, which flows on the $\eta=0$ bonds.
    The green and brown circular arrows denote loop currents
    with opposite chirality.
    (c) Obtained $n$ dependences of $J_1$ and $J_2$ for $\eta=0.01$ eV.
    (d) $\eta$ dependences of $J_1$ and $J_2$ at $n=2.5$.
    The red dashed line represents the line proportional to $\eta$.}
    \label{fig6}
\end{figure}

Next, we study the impurity effect in the single-$\boldsymbol{Q}$ cLC order.
The single-$\boldsymbol{Q}$ cLC order without the impurity is shown in Fig. \ref{fig6} (a).
In this case, the unit cell is extended to $2\times1$, which includes six sites.
The single-$\boldsymbol{Q}$ cLC breaks the local TRS, 
while $M_{\mathrm{orb}}$ and $\sigma_{xy}$ vanish identically 
because the global TRS is preserved in the absence of the impurity.

%%%%%%%%%%%%%%%%%%%%%%%%%%%%%%%%%%%%%%%%%%%%%%%%%%
%%%%%%%%%%%%%%%%%%%%%%%%%%%%%%%%%%%%%%%%%%%%%%%%%%

Figure \ref{fig6}(b) shows the schematic picture of 
the current pattern when $n$ is far from the VHS filling $n_{\rm VHS}$ (left panel) and $n\approx n_{\rm VHS}$ (right panel).
The numerical results are shown in Fig. \ref{fig6}(c) for $\eta=0.01\,\mathrm{eV}$.
In the single-$\boldsymbol{Q}$ cLC order,
$|J_{i,j}|=|J_1|$ for all $\eta\ne0$ bonds (red arrows)
and $|J_{i,j}|=|J_2|$ for all $\eta=0$ bonds (blue arrows).
In Fig. \ref{fig6}(c), the obtained $J_1$ ($J_2$) is shown by red (blue) lines, which is schematically shown by red (blue) arrows in Fig. \ref{fig6}(b).
Interestingly, $J_1$ exhibits a sign change as a function of $n$ in Fig. \ref{fig6}(c).
Therefore, the triangular current loops are generated when $n$ is far from $n_{\mathrm{VHS}}$, while the hexagonal current loops are generated for $n\approx n_{\rm VHS}$, as summarized in Fig. \ref{fig6}(b).
Both $J_1$ and $J_2$ are linear in $\eta$, as shown in Fig. \ref{fig6}(d).
In Fig. \ref{fig6}(d), both $|J_1|$ and $|J_2|$ are linear in $\eta$ for $\eta\lesssim0.01$, 
where the relation $|J_1| \gg |J_2|$ holds.
In the present range $(0\le\eta\le0.03)$, interesting nonlinear behavior is obtained only for $|J_1|$, 
which tends to saturate for $\eta\gtrsim0.01$.
Such nonlinear behavior of the charge currents, which will depend on the filling $n$, 
is an interesting future problem.

\subsection{Impurity effect in the single-$\boldsymbol{Q}$ cLC}
\begin{figure}[htb]
    \centering
    \includegraphics[width=85mm]{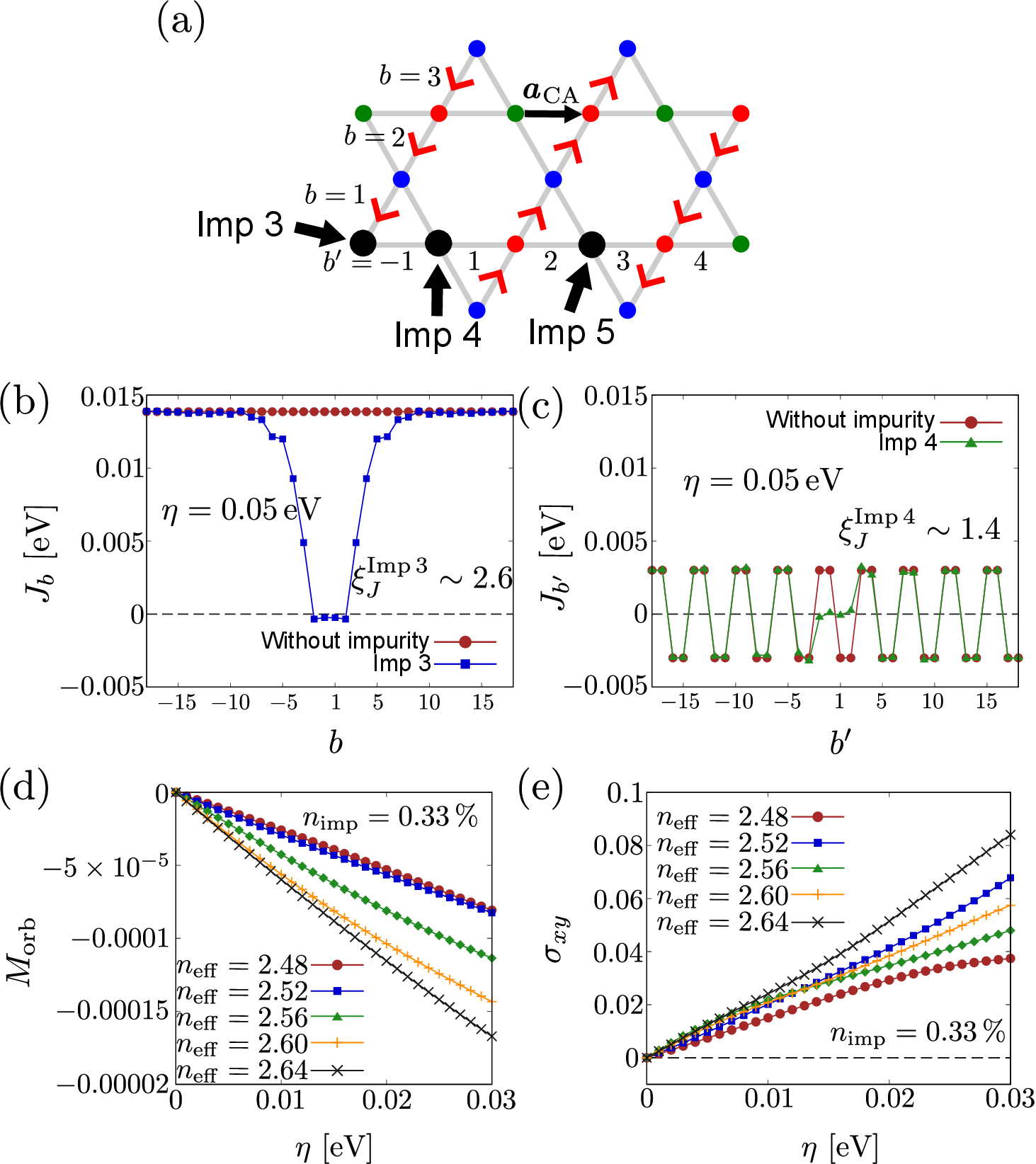}
    \caption{(a) Kagome lattice tight-binding model for 
    the single-$\boldsymbol{Q}$ cLC order and the position of 
    an introduced impurity in the Imp 3, Imp 4, and Imp 5 cases.
    The red arrows represent the direction of $\eta$. 
    Obtained currents on the bonds along a straight line through the impurity 
    in (b) the Imp 3 case and (c) the Imp 4 case for $\eta=0.05$ eV and $n_{\mathrm{eff}}=2.48$.
    Obtained $\eta$ dependences of (d) $M_{\mathrm{orb}}$ and 
    (e) $\sigma_{xy}$ under the condition of $n_{\mathrm{imp}}=0.33\%$ in the case of Imp 4.}
    \label{fig7}
\end{figure}

Next, we study the impurity effect in the single-$\boldsymbol{Q}$ cLC order 
in both the Imp 3 and Imp 4 cases shown in Fig. \ref{fig7}(a).
We first calculate the current distribution 
for $N=972$ with $M_x=9$ and $M_y=18$, which corresponds to $n_{\mathrm{imp}}=0.1\,\%$.
Figure \ref{fig7}(b) shows the obtained current on the $b$th bond $J_b$ in the case of Imp 3, 
which is inserted on the $\eta\ne0$ line.
The bond indices $b$ and $b'$ are shown in Fig. \ref{fig7}(a).
In the case of Imp 3, the suppressed $J_b$ near the impurity site recovers to the original value for $|b|>\xi_J^{\mathrm{Imp}\,3}\approx2.6$.
Figure \ref{fig7}(c) shows the bond current in the case of Imp 4, which is introduced on the line with $\eta=0$.
In this case, the correlation length $\xi_J^{\mathrm{Imp}\,4}$ is about $1.4$.

%%%%%%%%%%%%%%%%%%%%%%%%%%%%%%%%%%%%%%%%%%%%%
%%%%%%%%%%%%%%%%%%%%%%%%%%%%%%%%%%%%%%%%%%%%%

Next, we calculate $M_{\mathrm{orb}}$ and $\sigma_{xy}$ 
in the single-$\boldsymbol{Q}$ cLC order with an impurity.
In the case of Imp 3, both $M_{\mathrm{orb}}$ and $\sigma_{xy}$ vanish because the global TRS is preserved.
Therefore, we consider the Imp 4 case by setting $N=300$ with $M_x=5$ and $M_y=10$, 
which corresponds to $n_{\mathrm{imp}}=0.33\,\%$.
In the Imp 3 case, the TRS operation does not change the electronic state, 
with the consideration of a $\pi$ rotation and a global shift of $\pm2\boldsymbol{a}_{\mathrm{CA}}$.
In contrast, in the Imp 4 and Imp 5 cases, the TRS operation alters each electronic state, 
even if any rotations and global shifts are considered.
Therefore, while Imp 3 does not globally break the TRS, both Imp 4 and Imp 5 break the global TRS.
Figures \ref{fig7}(d) and 7(e) represent the $\eta$ and $n_{\mathrm{eff}}$ dependences of 
$M_{\mathrm{orb}}$ and $\sigma_{xy}$, respectively, in the case of Imp 4.
(The obtained $M_{\mathrm{orb}}$ and $\sigma_{xy}$ are much smaller than those for the triple-$\boldsymbol{Q}$ order in Figs. \ref{fig4} and  \ref{fig5}.)
Interestingly, a single nonmagnetic impurity gives rise to finite $M_{\mathrm{orb}}$ 
in the single-$\boldsymbol{Q}$ cLC state, where $M_{\mathrm{orb}}$ is absent without the impurity.
We note that the impurity-induced $M_{\mathrm{orb}}$ and $\sigma_{xy}$ due to Imp 5 
in Fig. \ref{fig7}(a) are the inverses of those in Figs. \ref{fig7}(d) and 7(e).
Nonetheless, impurity-induced $M_{\mathrm{orb}}$ and $\sigma_{xy}$ obtained 
in Figs. \ref{fig7}(d) and 7(e) can be observed in microscopic measurements.

\section{Summary}
In this paper, we analyzed the giant unit-cell ($N\le1200$) kagome lattice model with single impurity potential to understand the impurity effects on the cLC electronic states accurately.
The loop current is found to be strongly suppressed 
within the current correlation length $\xi_J$ centered on the impurity site,
where $\xi_J$ is inversely proportional to the cLC order parameter $\eta$.
For example, we obtain 
$\xi_J^{\mathrm{Imp}\,1}\approx3.2$ and $\xi_J^{\mathrm{Imp}\,2}\approx2.5$ for $\eta=0.05$ eV, while 
we obtain $\xi_J^{\mathrm{Imp}\,1}\approx4.8$ and $\xi_J^{\mathrm{Imp}\,2}\approx4.1$ for $\eta=0.01$ eV.
Thus, the obtained $\xi_J$ increases as $|\eta|\rightarrow0$.
This result reminds us of the BCS coherence length 
$\xi_{\rm BCS}=\pi v_{\rm F}/\Delta_{\rm SC}$.
Under the triple-$\boldsymbol{Q}$ cLC order, 
the folded Fermi surfaces meet around the $\Gamma$ point in the folded BZ.
The band-hybridization gap $\Delta_{\mathrm{cLC}}$ due to the cLC order will be proportional to $\eta$.
Therefore, the relationship in which $\xi_J$ increases as $|\eta|\rightarrow0$ 
is naturally expected in kagome metals.

\begin{figure}[htb]
    \centering
    \includegraphics[width=85mm]{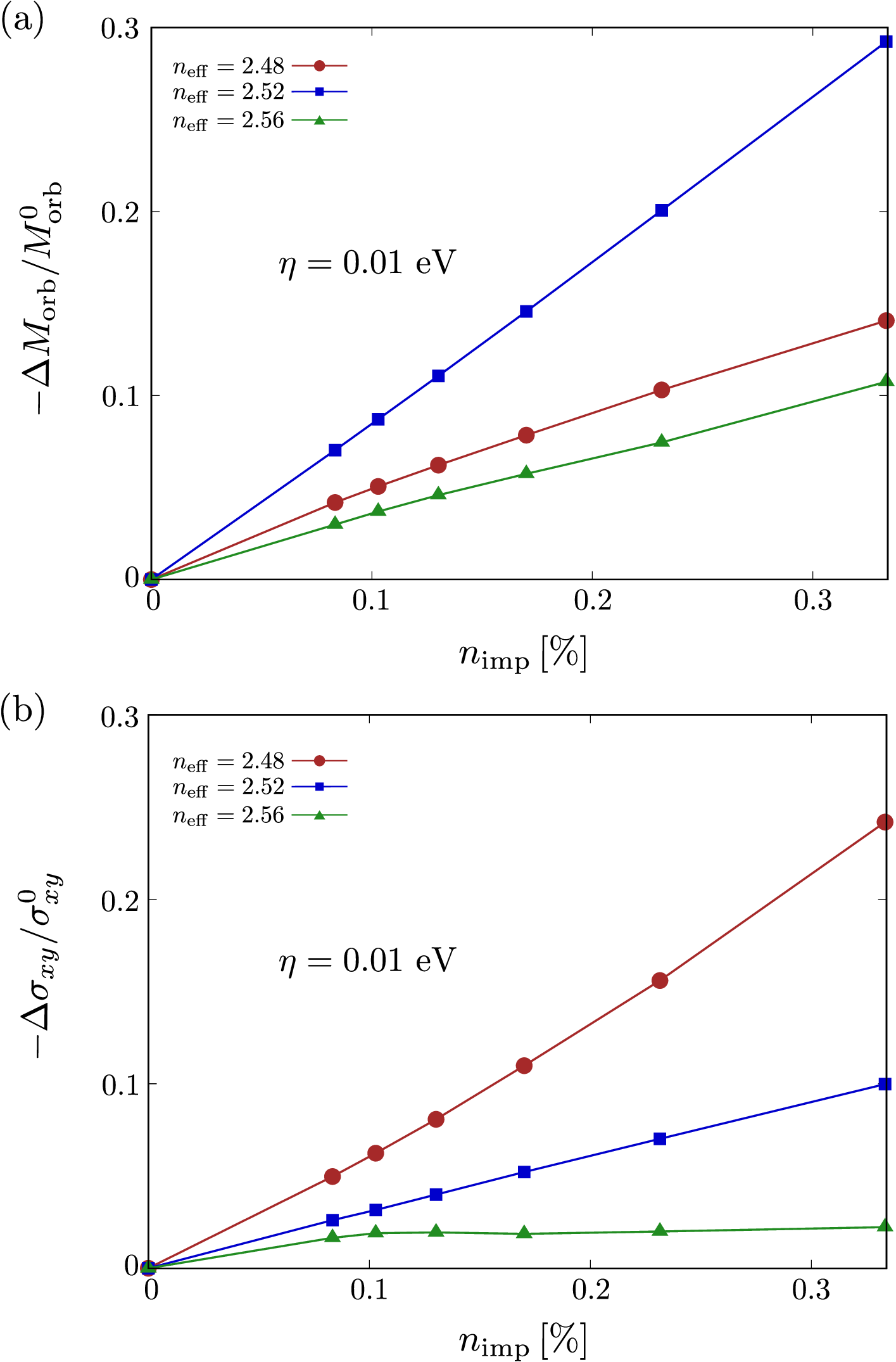}
    \caption{$n_{\mathrm{imp}}$ dependences of $-\Delta X/X^0$ 
    ($X^0$ denotes quantity without the impurity) 
    for (a) $X=M_{\mathrm{orb}}$ and (b) $X=\sigma_{xy}$ 
    derived from $M_x=M_y=5$-10 models at 
    $n_{\mathrm{imp}}=$2.48, 2.52, and 2.56.
    We set $\eta=0.01$ eV.}
    \label{fig8}
\end{figure}

In addition, we found that both the uniform orbital magnetization $M_{\rm orb}$ and 
the anomalous Hall conductivity $\sigma_{xy}$ are drastically suppressed by dilute impurities.
To confirm this statement, we perform a numerical study for $M_x=M_y=5$-10 models in Fig. 8, 
which shows the $n_{\mathrm{imp}}$ dependence of the suppression ratio $-\Delta X/X^0$
for $X=M_{\mathrm{orb}}$  [Fig. \ref{fig8}(a)] and $X=\sigma_{xy}$ [Fig. \ref{fig8}(b)].
Here, the suppression ratios for both $M_{\mathrm{orb}}$ and $\sigma_{xy}$ are linear in $n_{\mathrm{imp}}$ in the low-impurity region, 
except $X=\sigma_{xy}$ at $n_{\mathrm{eff}}=2.56$ 
may be due to 
the proximity to the Van Hove filling.
The origin of this exceptional behavior at $n_{\mathrm{eff}}=2.56$ is left for future work.

Therefore, the results in Figs. 4 and 5 have sufficient calculation accuracy.

For both $X=M_{\rm orb}$ and $\sigma_{xy}$, the suppression ratio $R=-\Delta X/X^0$ 
can exceed $50\%$ with the introduction of just $1\%$ impurities for $\eta<0.01\,\mathrm{eV}$;
see Fig. \ref{fig5} (c) for $n_{\rm imp}=0.33\,\%$.
The obtained giant impurity effects on $M_{\rm orb}$ and $\sigma_{xy}$ in the $2\times2$ cLC phase 
provide a natural explanation of why the cLC electronic states in kagome metals are sensitive 
to the dilute impurities, 
as indicated by the $n_{\mathrm{imp}}$-sensitive chiral QPI signal in the STM measurements 
\cite{BO1,QPI,QPI2}.

%%%%%%%%%%%%%%%%%%%%%%%%%%%%%%%%%%%%%%%%%%%%%
%%%%%%%%%%%%%%%%%%%%%%%%%%%%%%%%%%%%%%%%%%%%%

Recently, Asaba {\it et al.} discovered that the
single-$\boldsymbol{Q}$ cLC order emerges above $T_{\rm BO}=90$ K
based on the magnetic torque measurement 
\cite{torque}.
In the single-$\boldsymbol{Q}$ cLC state,
both $M_{\rm orb}$ and $\sigma_{xy}$ disappear 
because the global TRS is preserved.
However, we find that an impurity in a nanoscale cluster model 
gives rise to finite $M_{\rm orb}$ and $\sigma_{xy}$,
although they vanish if we take the average of the impurity sites.
Interestingly, the single-$\boldsymbol{Q}$ cLC orders induce nontrivial loop current patterns, 
so the emerging characteristic magnetic fields can be observed experimentally.

In the present study,
we assumed that the cLC order parameter $\eta$ is constant for all the nearest bonds.
However, $\eta$ would be suppressed near the impurity site because the cLC order parameter 
is not an $s$ wave, which is usually fragile against impurity scattering.
In this respect, the drastic impurity effects on $M_{\rm orb}$ and $\sigma_{xy}$ obtained here will be underestimated.
This is an important future problem for studying the impurity effect on $\eta$ based on a microscopic theory.

%acknowledge
\begin{acknowledgments}
We are grateful to Y. Matsuda, T. Shibauchi, K. Hashimoto, T. Asaba, and S. Suetsugu for very useful discussions.
This study was supported by Grants-in-Aid for Scientific Research from MEXT of Japan (Grants No. JP24K00568, No. JP24K06938, No. JP23K03299, and No. JP22K14003).
\end{acknowledgments}

%%%%%%%%%%%%%%%%%%%%%%%%%%%%%%%%%%%%%%%%%%%%%%%%%%%%%%%

\end{document}